\documentclass[a4paper,fleqn,usenatbib]{mnras}

\usepackage[T1]{fontenc}
\usepackage{ae,aecompl}

\usepackage{graphicx}	
\usepackage{amsmath}	
\usepackage{amssymb}	

\title[Repulsive force]{A repulsive force in the Einstein theory}

\author[N. Gorkavyi and A. Vasilkov]{
Nick Gorkavyi$^{1}$\thanks{E-mail: gorkavyi@gist.us}
and Alexander Vasilkov$^{1}$
\\
$^{1}$Science Systems and Applications, Inc., 10210 Greenbelt Rd.,
Lanham, MD 20706, USA
}

\date{Accepted 2016 June 20. Received 2016 June 20; in original form 2016 April 28}

\pubyear{2016}

\begin{document}
\label{firstpage}
\pagerange{\pageref{firstpage}--\pageref{lastpage}}
\maketitle

\begin{abstract}
The Laser Interferometer Gravitational-Wave Observatory (LIGO) detection of gravitational waves that take away 5\% of the total mass of 
two merging black holes points out on the importance of considering varying
gravitational mass of a system in the framework of the Einstein general theory of relativity.
We calculate the acceleration of a particle in the non-stationary field of a quasi-spherical system
composed of a large number of objects emitting gravitational waves. 
It is shown that reduction of the gravitational mass of the system 
due to emitting gravitational waves leads to a repulsive gravitational force
that diminishes with time but never disappears. This repulsive force may be
related to the observed expansion of the Universe.
\end{abstract}




\section{Introduction}

Recently the Laser Interferometer Gravitational-Wave Observatory (LIGO) Scientific Collaboration announced a detection of gravitational waves
caused by the merger of two black holes with masses of about 36 and 29 times the mass of the Sun \citep{Abbot}.
About three times the mass of the Sun, i.e. about 5\% of the initial total mass of the black holes, 
was converted into gravitational waves. The LIGO detection of the gravitational waves
confirms a main prediction of Einstein's general theory of relativity. 
The merger of black holes is currently rare, however the collisions and merger of black
holes will be a common event at the final stage of collapsing Universe \citep{Sikkema,Banks}. 
Let us estimate a fraction of black hole masses that converts into gravitational waves
using a model of collapsing Universe called the `big crunch'.
Currently the Universe composed of approximately 10$^{22}$ or 2$^{73}$ stars.
Assuming that all stars are converted into black holes,
we can consider 73 binary mergers of the black holes in the `big crunch'
with 5\% mass of every collision of the black holes transformed to
gravitational waves. A final black hole will have a mass of 0.95$^{73} \sim 0.02$
of the initial mass of the Universe. The remaining mass of the Universe (98\%) will
be converted into gravitational waves.

How the total gravitational mass of the Universe changes when a significant
fraction of the mass of the black holes is converted into gravitational waves? 
An answer to this question depends on treating the gravitational field sources 
(i.e. the right part of Einstein's equations) in the presence of gravitational waves.
At the earlier stage of the development of the general theory of relativity (years 1913-1916)
Einstein supposed that the energy of matter and the energy of gravitational field are 
equivalent as a source of gravitational field and included the gravitational energy
in the right part of his equations. Later, after discussions with Schr\"odinger and other scientists about
the non-tensor and non-local nature of the energy-impulse of the gravitational field,
Einstein changed his mind. Since 1917 he never included the gravitational energy in
the right part of his equations and pointed out that a single source of gravitational field is
the energy-impulse tensor of ordinary matter and electromagnetic field: `$T_{ik}$ represents
the energy which generates the gravitational field, but is itself of non-gravitational character,
as for example the energy of the electromagnetic field, of the density of 
ponderable matter etc' \citep{Einstein53}. This Einstein's point of view was shared by \citet{Schrodinger}, \citet{Eddington} and  \citet{Chandrasekhar}.
In addition, \citet{Schrodinger} pointed out that the total mass of the Universe can change
when the Universe expands.

There is an opinion that it is possible to expand the Ricchi tensor (the left part of
Einstein's equations) into linear and non-linear parts, then to move the non-linear part
of the Ricchi tensor to the right part of the equations, to call it as a pseudo-tensor
of energy-impulse of the gravitational field (or/and gravitational waves), and to treat it
as an additional source of the gravitational field \citep{Weinberg}. However, \citet{Misner} 
showed that such a procedure resulted in a non-Einstein theory of gravitation.  
Such a contradiction was considered to be insignificant at that time because
the effect of emitting gravitational waves on the total mass of a system was
supposed to be negligible. However, the LIGO discovery proved non-negligible loss of
the total mass of two black holes due to gravitational waves and made a question
about the contribution of gravitational waves to the total mass of a system be an urgent problem.

In this paper we consider a model of the universe with decreasing gravitational mass in the
framework of the general theory of relativity. It should be noted that the model of
the universe with varying gravitational mass is not quite new. For example, \citet{Tolman} 
admitted that the total gravitational mass of the universe changes due to transformation of matter to
gravitational waves. A model of the universe with growing gravitational mass was
considered by \citet{Hoyle}. \citet{Kutschera} considered a model of relativistic fireball 
with variable gravitational mass. 

We suppose that gravitational waves do not contribute to 
the total mass of a system. We show that reduction of the gravitational mass of the system
due to emitting gravitational waves leads to a repulsive gravitational force.
The observed expansion of the Universe suggests that the repulsive forces were
extremely large in the past. \citet{Guth} proposed that the Universe in the early stage of
its history experienced the inflation caused by a scalar field called `inflaton'.
Recent observations showed that the Universe currently expands with acceleration
\citep{Riess,Perlmutter}. To explain the accelerating expansion of the Universe
there were developed numerous non-Einstein's theories (see e.g. \citet{Bamba}) and
cosmological models with hypothetical `dark energy'. However this does not succeed in 
a decent solution of the problem of the accelerating expansion of the Universe - 
see a review by \citet{Joyce}.

We show that a repulsive gravitational force can be derived in the framework of the
classic general theory of relativity. The repulsive terms in the Einstein theory have been considered by \citet{Hilbert},  \citet{McVittie}, \citet{Zeldovich} and \citet{McGruder}. It should be noted that \citet{Hilbert} was the first
who showed that the equation of acceleration of a probe particle in the general theory of
relativity contained two terms which represented both attractive and repulsive forces. The existence of the attractive and 
repulsive forces in the general theory of relativity has been discussed for about a century - 
see a review by \citet{McGruder}. We derive a general expression for the acceleration of 
a probe particle in a given diagonal metric that contains repulsive terms. Our analysis is
carried out for the case of weak gravitational field and non-relativistic velocity of a probe particle.
We show that the cosmic repulsive force can be explained in the classic general theory of relativity 
without additional hypotheses.
  
\section{Acceleration of a particle in gravitational field}


\subsection{General equation of gravitational acceleration}
\label{sec:maths} 

In the general theory of relativity the four-dimensional acceleration of a probe particle
is defined by the equation of the geodesic curve (see, e.g., \citet*{Weinberg}):
\begin{equation}
\frac{d^{2}x^{\mu}}{d{\tau}^2}=-{{\Gamma}^{\mu}_{{\nu}{\lambda}}}{\frac{dx^{\nu}}{d{\tau}}}{\frac{dx^{\lambda}}{d{\tau}}}, 
\label{eq:eq1}
\end{equation}
where $\tau$ is the time in the coordinate system bound with the particle. Greek indices run the values of 0,1,2,3.
Equation~(\ref{eq:eq1}) is written in the system where the light velocity is assumed to be $c=1$.
The time coordinate is $x^0=t$, the radial coordinate in the spherical system is $x^1=r$.
The three-dimensional acceleration, that is measured by a distant observer, is expressed as follows:
\begin{equation}
\frac{d^{2}x^i}{dt^2}=-{{\Gamma}^i_{{\nu}{\lambda}}}{\frac{dx^{\nu}}{dt}}{\frac{dx^{\lambda}}{dt}}+{{\Gamma}^0_{{\nu}{\lambda}}}{\frac{dx^{\nu}}{dt}}{\frac{dx^{\lambda}}{dt}}{\frac{dx^i}{dt}}. 
\label{eq:eq2}
\end{equation}
Here the Latin indices run the values 1,2,3 and are related to space coordinates only. 
Equation~(\ref{eq:eq2}) can be rewritten in the form \citep{Weinberg}:
\begin{eqnarray}
\frac{d^{2}x^i}{dt^2}=-{\Gamma}^i_{00}-2{{\Gamma}^i_{0j}}{\frac{dx^j}{dt}}
-{{\Gamma}^i_{jk}}{\frac{dx^j}{dt}}{\frac{dx^k}{dt}}+ \nonumber \\
({\Gamma}^0_{00}+2{{\Gamma}^0_{0j}}{\frac{dx^j}{dt}}+ {{\Gamma}^0_{jk}}{\frac{dx^j}{dt}}{\frac{dx^k}{dt}}){\frac{dx^i}{dt}}. 
\label{eq:eq3}
\end{eqnarray}
The Christoffel symbols are expressed through components of the metric tensor:
\begin{equation}
{{\Gamma}^{\sigma}_{{\mu}{\lambda}}}={\frac{1}{2}}{{g}^{{\nu}{\sigma}}}({\frac{\partial{g_{{\mu}{\nu}}}}{\partial{x^{\lambda}}}}+{\frac{\partial{g_{{\lambda}{\nu}}}}{\partial{x^{\mu}}}}-{\frac{\partial{g_{{\mu}{\lambda}}}}{\partial{x^{\nu}}}}).
\label{eq:eq4}
\end{equation}
Let us restrict our consideration by radial movements only and write all non-zero
components of the Christoffel symbols for non-stationary diagonal metrics where the component $g_{11}$ does not depend on time and $g^{00}=1/g_{00}$ and $g^{11}=1/g_{11}$:
\begin{equation}
\Gamma^0_{00}={\frac{1}{2g_{00}}}{\frac{\partial{g}_{00}}{\partial{t}}},
\label{eq:eq5}
\end{equation}
\begin{equation}
\Gamma^1_{00}=-{\frac{1}{2g_{11}}}{\frac{\partial{g}_{00}}{\partial{r}}},
\label{eq:eq6}
\end{equation}
\begin{equation}
\Gamma^1_{11}={\frac{1}{2g_{11}}}{\frac{\partial{g}_{11}}{\partial{r}}},
\label{eq:eq7}
\end{equation}
\begin{equation}
\Gamma^0_{01}={\frac{1}{2g_{00}}}{\frac{\partial{g}_{00}}{\partial{r}}}.
\label{eq:eq8}
\end{equation}
From equation~(\ref{eq:eq3}) we get accounting for all non-zero components of the Christoffel symbols:
\begin{equation}
\frac{d^{2}r}{dt^2}=-{\Gamma}^1_{00}-{{\Gamma}^1_{11}}{(\frac{dr}{dt})^2}
+({\Gamma}^0_{00}+2{{\Gamma}^0_{01}}{\frac{dr}{dt}}){\frac{dr}{dt}}.
\label{eq:eq9}
\end{equation}
Taking into account equations~(\ref{eq:eq5})-(\ref{eq:eq8}) we get:
\begin{eqnarray}
\frac{d^{2}r}{dt^2}={\frac{1}{2g_{11}}}[{\frac{\partial{{g}_{00}}}{\partial{r}}}-{\frac{\partial{{g}_{11}}}{\partial{r}}}{(\frac{dr}{dt})^2}]+
\nonumber\\
{\frac{1}{2g_{00}}}({\frac{\partial{{g}_{00}}}{\partial{t}}}+2{\frac{\partial{{g}_{00}}}{\partial{r}}}{\frac{dr}{dt}}){\frac{dr}{dt}}.
\label{eq:eq10}
\end{eqnarray}
Equation~(\ref{eq:eq10}) can be rewritten in the form of a polynomial with respect to the velocity:
\begin{eqnarray}
a={\frac{1}{2g_{11}}}{\frac{\partial{{g}_{00}}}{\partial{r}}}
+{\frac{1}{2g_{00}}}{\frac{\partial{{g}_{00}}}{\partial{t}}}v+({\frac{1}{g_{00}}}{\frac{\partial{{g}_{00}}}{\partial{r}}}
-{\frac{1}{2g_{11}}}{\frac{\partial{{g}_{11}}}{\partial{r}}})v^2.
\label{eq:eq11}
\end{eqnarray}
Equation~(\ref{eq:eq11}) is a general expression of the radial three-dimensional 
gravitational acceleration in the Einstein theory for non-stationary diagonal metrics with the component $g_{11}$ that does not depend on time.

\subsection{Gravitational acceleration in the Schwarzschild metric}

Let us write components of the metric tensor for the stationary Schwarzschild metric:

\begin{equation}
\ {g}_{00}=-(1-{\frac{r_0}{r}}),
\label{eq:eq12}
\end{equation}
\begin{equation}
\ {g}_{11}=(1-{\frac{r_0}{r}})^{-1},
\label{eq:eq13}
\end{equation}
where $r_0$ is the Schwarzschild radius. Accounting for equations~(\ref{eq:eq12}) and (\ref{eq:eq13})
we get equation~(\ref{eq:eq11}) of the gravitational acceleration in the form:
\begin{equation}
\ a=-{\frac{r_0}{2r^2}}(1-{\frac{r_0}{r}})+{\frac{3r_0}{2r^2}}{v^2}{(1-{\frac{r_0}{r}})}^{-1}.
\label{eq:eq14}
\end{equation}
Equation~(\ref{eq:eq14}) was published by \citet{Hilbert} who    
pointed out that this equation contains a repulsive term along with the Newton
attraction term. Based on equation~(\ref{eq:eq14}) \citet{McVittie} concluded that
a central body can cause the repulsive force in the Einstein gravitation.
\citet{Zeldovich} considered the repulsive force be fictitious because it caused by
time deceleration in the vicinity of a black hole. Indeed, a body dropping on the black hole
cannot reach up the Schwarzschild radius from the point of view of a distant observer.
This means that the body moving to the black hole is decelerated. The deceleration of the body
can be attributed to "a repulsive force". The higher is the body velocity, the larger should 
be the repulsive force, thus supporting the velocity squared term in equation~(\ref{eq:eq14}).
\citet{McGruder} reviewed all discussions about equation~(\ref{eq:eq14}) and pointed out that 
the repulsive force still existed even at large distances from the black hole for
sufficiently high  velocities of the probe body.

Here and elsewhere we will consider weak gravitational fields, that is we assume distances
much larger than the Schwarzschild radius $r >> r_0$, and low velocities of a probe body $v << 1$.
In this approximation, equation~(\ref{eq:eq14}) describes the ordinary Newton gravity.


\section{Metric of an object with varying mass}

Let us consider a quasi-spherical system composed of a large number of objects
emitting gravitational waves such as merging black holes discovered by the LIGO.
At large distances from the system the gravitational field can be treated as a weak one.
For such a weak gravitational field we have:
\begin{equation}
\ {g}_{{\mu}{\nu}}={\eta}_{{\mu}{\nu}}+{h}_{{\mu}{\nu}},
\label{eq:eq15}
\end{equation}
where $\eta_{\mu \nu}$ is the Minkovsky tensor and the inequality 
$\eta_{\mu \nu} >> h_{\mu \nu}$ is valid for all components.

Beginning with \citet{Einstein18}, it has been shown that the Einstein equations (in the unit system where the light velocity is not equal to unity)
\begin{equation}
\ {R}_{{\mu}{\nu}}=-{\frac{8{\pi}G}{c^4}}({T}_{{\mu}{\nu}}-{\frac{1}{2}}{g}_{{\mu}{\nu}}{T}^{\lambda}_{\lambda})
\label{eq:eq16}
\end{equation}
transform to an analogy of the wave equation for weak gravitational fields. Following
\citet{Weinberg} and \citet{Landau} we can write this equation as follows:
\begin{equation}
\ {\square}^2{h}_{{\mu}{\nu}}=-{\frac{16{\pi}G}{c^4}}{S}_{{\mu}{\nu}},
\label{eq:eq17}
\end{equation}
where 
\begin{equation}
\ {S}_{{\mu}{\nu}}\equiv {T}_{{\mu}{\nu}}-{\frac{1}{2}}{\eta}_{{\mu}{\nu}}{T}^{\lambda}_{\lambda}.
\label{eq:eq18}
\end{equation}
A solution of equation~(\ref{eq:eq17}) is the following expression with the retarded potentials
\citep{Weinberg,Landau}:
\begin{equation}
\ {h}_{{\mu}{\nu}}(r,t)={\frac{4G}{c^4}}{\int}{\frac{{S}_{{\mu}{\nu}}(t-r/c)}{r}}dV.
\label{eq:eq19}
\end{equation}
Assuming sufficiently low velocities of objects in the quasi-spherical system
we can get for the zero component \citep{Weinberg}
\begin{equation}
\ {S}_{00}(t-r/c)={\frac{1}{2}}{T}_{00}(t-r/c)\approx {\frac{1}{2}}c^2 {\rho}(t-r/c).
\label{eq:eq20}
\end{equation}
If we assume that the mass of the system is concentrated at small radii, we can write:
\begin{equation}
\ {M}(t-r/c)={\int}{\rho}(t-r/c)dV.
\label{eq:eq21}
\end{equation}
The quasi-spherical system composed of objects which are losing their mass by radiating
gravitational waves. This leads to decreasing the total mass of the system, $M$, and its density, $\rho $.
Change of the mass causes the external metric of the quasi-spherical system be non-stationary.
It should be noted that the change of the gravitational mass of the system does not violate
the Birkhoff theorem because the Birkhoff theorem is valid for spherically-symmetric stationary fields,
for example for the Schwarzschild metric. The existence of gravitational radiation from merging
black holes in the system under consideration violates the spherical symmetry of the system thus
making the Birkhoff theorem inapplicable to the system, as it has been noted by \citet{Synge}.
Therefore, our non-stationary metric is substantially different from the Schwarzschild metric.
\citet{Zeldovich} pointed out that a binary star system, which is losing its mass due to gravitational waves,
has a spherically symmetric monopole component of
gravitational field that is proportional to the varying mass. \citet{Kutschera} found a solution for 
monopole gravitational waves from relativistic fireballs with variable gravitational mass. 
In the paper we consider this spherically-symmetric monopole component of
gravitational field of an object with a diminishing mass.

We showed that the Einstein equations have an exact solution with a variable mass of the system 
in the approximation of weak gravitational fields. The solution does not violate the covariant
analogy of the mass conservation law and the equality of the covariant derivative of the energy-impulse tensor
(describing a source of the gravitational field) to zero:   
\begin{equation}
\ {{T}^{{\mu}{\nu}}}_{;\mu} \equiv {\frac{\partial{T^{{\mu}{\nu}}}}{\partial{x^{\mu}}}}
+{{\Gamma}^{\mu}_{{\mu}{\lambda}}}{T}^{{\lambda}{\nu}}+{{\Gamma}^{\nu}_{{\mu}{\lambda}}}{T}^{{\mu}{\lambda}}=0.
\label{eq:eq22}
\end{equation}
Because of non-zero Christoffel symbols equation~(\ref{eq:eq22}) allows variability of the gravitational mass
of the system. Equation~(\ref{eq:eq22}) can be used to derive an expression for decreasing energy
of a rotating bar caused by radiating gravitational waves \citep{Eddington}.
The LIGO Collaboration used the Einstein equations to calculate gravitational waves \citep{Abbot}.
This proves that decreasing gravitational mass of the merging black holes does not contradict
the Einstein equations. It is obvious that the Einstein equations can be also applied to a system of
multiple merging black holes, which is the case under consideration in our paper.
Therefore, the decreasing gravitational mass agrees with both the Einstein equations and 
the conservation law equation~(\ref{eq:eq22}) in the case when the decreasing gravitational mass
is related to perturbations of the space similar to gravitational waves.

Using  equations~(\ref{eq:eq19})-(\ref{eq:eq21}) we get for the zero component, $h_{00}$,
the following expression:
\begin{equation}
\ {h}_{00}=\frac{2GM(t-r/c)}{rc^2}.
\label{eq:eq23}
\end{equation}
A similar solution was found by \citet{Kutschera} for the perturbation of the Schwarzschild metric 
due to changing gravitational mass of relativistic fireballs. 
Putting equation~(\ref{eq:eq23}) into equation~(\ref{eq:eq15}) we get an expression for 
the zero component of the metric tensor:  
\begin{equation}
\ {g}_{00}=-[1-\frac{2GM(t-r/c)}{rc^2}].
\label{eq:eq24}
\end{equation}
Equation~(\ref{eq:eq24}) formally resembles the corresponding zero component of the
Schwarzschild metric of equation~(\ref{eq:eq12}).
However, the metric of equation~(\ref{eq:eq24}) is essentially different from the Schwarzschild metric
because it describes the varying gravitational mass.

Assuming weak gravitational fields and low velocities and using equation~(\ref{eq:eq11}) we get the following
expression for the gravitational acceleration:
\begin{equation}
\ a \approx {\frac{c^2}{2}}{\frac{\partial{{g}_{00}}}{\partial{r}}}={\frac{\partial}{\partial{r}}}
{\frac{GM(t-r/c)}{r}}
\label{eq:eq25}
\end{equation}
or:
\begin{equation}
\ a \approx - {\frac{GM(t-r/c)}{r^2}}+{\frac{G}{r}}{\frac{\partial{M(t-r/c)}}{\partial{r}}}.
\label{eq:eq26}
\end{equation}
The second term in equation~(\ref{eq:eq26}) originates from the variability of the gravitational mass
and the fact of a finite velocity of the mass perturbation spreading.

\section{Physical interpretation of the solution} 

\begin{figure}
    \includegraphics[width=\columnwidth]{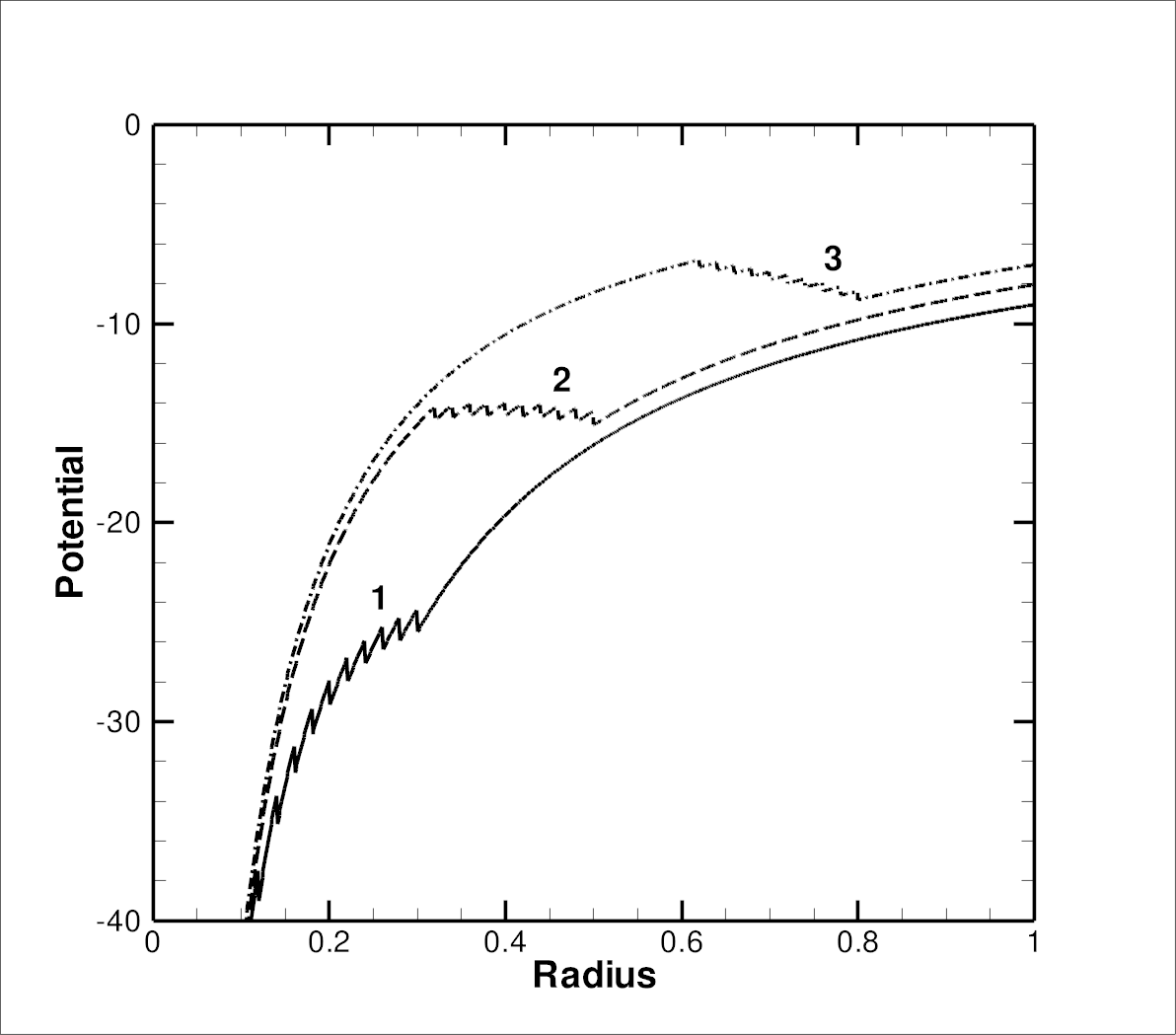}
    \caption{Gravitational potential (in $10^{-21}$ cm$^2$/s$^2$) as a function of the radius
(in billion light years) for different times. The curves 1,2,3 correspond to times
from the start of changing mass of the system: $t=0.3, 0.5, 0.8$ billion years. For better visualization, the curve 2 shifted on 1 unit versus the curve 3; the curve 1 shifted on 2 units versus the curve 3}
    \label{fig:fg1} 
\end{figure}

To clarify the physical sense of equation~(\ref{eq:eq26}) we rewrite the gravitational acceleration 
(\ref{eq:eq25}) through a potential, $\phi$:
\begin{equation}
\ a =-\frac{{\partial}{\phi}}{\partial{r}},
\label{eq:eq27}
\end{equation}
where
\begin{equation}
\ {\phi}=-{\frac{GM(t-r/c)}{r}}.
\label{eq:eq28}
\end{equation}
Let us consider the potential (equation~\ref{eq:eq28}) assuming that the model Universe with ${M_0}={10^{56}}$ g (common estimation of the baryon mass) composed of black holes
has underwent 10 mergers of the black holes with a time interval of 20 Myr.
Every merger of the black holes results in a 5\% loss of the system mass. The final mass of the system after 10 mergers of all black holes will be equal to ${(1-0.05)^{10}}\approx 0.60$. This 40\% mass reduction causes a wave of changing potential
that expands into a region of large radii. Fig.1~\ref{fig:fg1} shows the gravitational
potential as a function of radius for a few values of time since the beginning of the change of
mass. If the mergers of the black holes occur more frequent, the curves in Fig.~\ref{fig:fg1}
become smoother.

The gravitational acceleration is always directed against the gradient of the gravitational
potential. That is why the curves of the gravitational potential in Fig.~\ref{fig:fg1}
show the direction of acceleration. It can be seen in Fig.~\ref{fig:fg1} that a wave of
changing potential transforms into a wave of the repulsive force in a some period of time.

Let us describe the diminishing mass of the model Universe by an exponential function, $exp(-\alpha t)$,
where $\alpha$ characterizes the speed of the mass reduction. If the change of the central gravitational
mass follows the expression:
\begin{equation}
\ M={M_0}{\exp{[-{\alpha}(t-r/c)}]},
\label{eq:eq29}
\end{equation}
then equation~(\ref{eq:eq26}) can be written in the form:
\begin{equation}
\ a = -{\frac{GM}{r^2}}+{\frac{\alpha}{c}}{\frac{GM}{r}}.
\label{eq:eq30}
\end{equation}
If $\alpha > 0$ the second term of equation~(\ref{eq:eq30}) is positive and describes
the repulsive force. The faster decrease of the mass, i.e. the larger $\alpha$, corresponds to 
the stronger repulsive force.

\begin{figure}
    \includegraphics[width=\columnwidth]{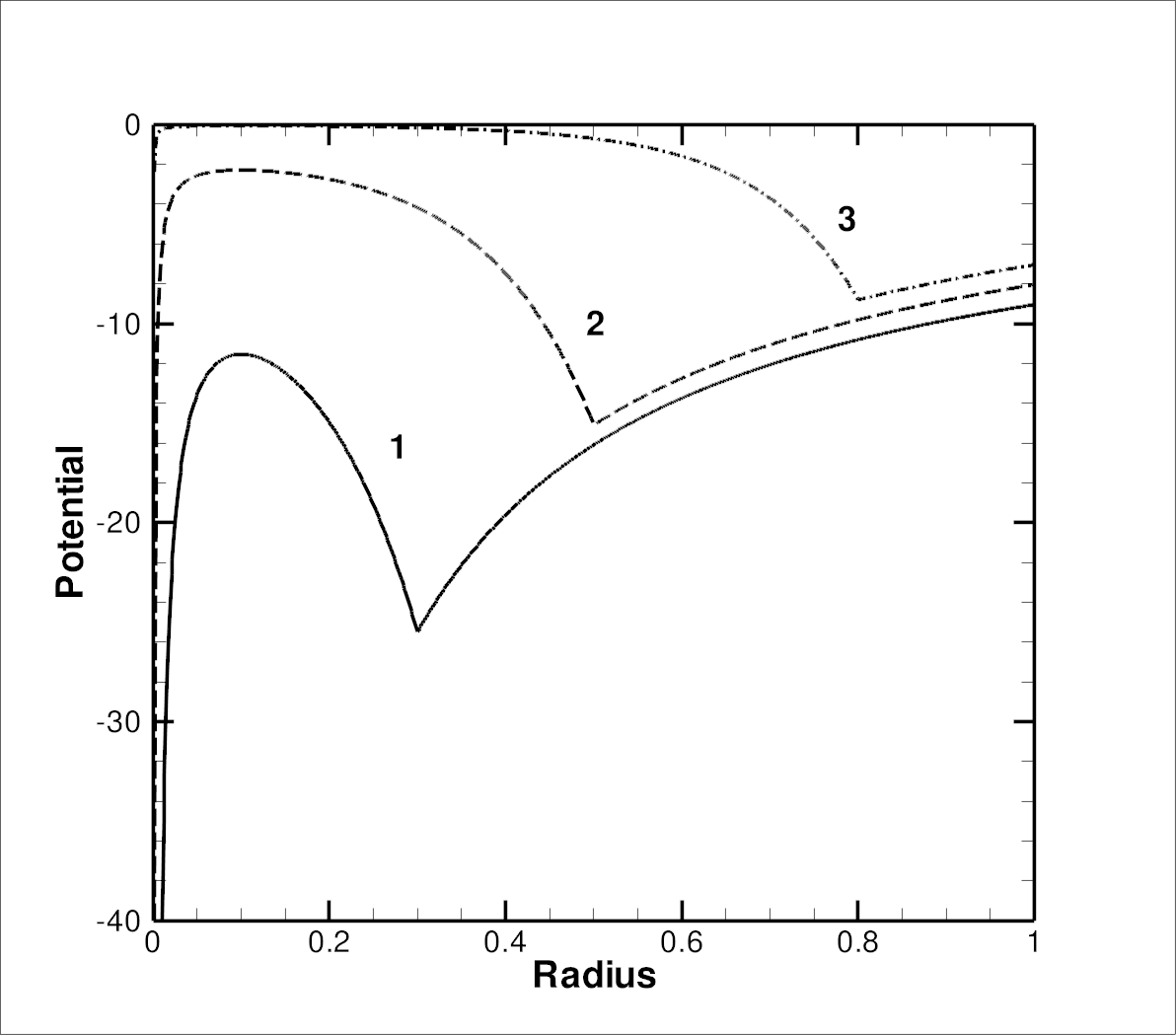}
    \caption{Similar to Fig.~\ref{fig:fg1} but for the exponential decrease of
the mass with $\alpha=10$}
    \label{fig:fg2} 
\end{figure}

In the case of the exponential decrease of the mass equation~(\ref{eq:eq29}) the gravitational potential
is expressed as:
\begin{equation}
\ \phi=-{\frac{GM_0}{r}}{\exp{[-{\alpha}(t-r/c)}]}.
\label{eq:eq31}
\end{equation}
The gravitational potential of equation~(\ref{eq:eq31}) is shown in Fig.~\ref{fig:fg2} as
a function of the radius for three values of time. In calculation of the gravitational potential
we assume that the mass of the system is constant till a some moment of time then it begins
to decrease in accordance with the exponential function. For radii $r > ct$, the gravitational potential 
keeps its shape. At $r < ct$ the exponent in equation~(\ref{eq:eq31}) becomes negative. The gradient of the potential
changes its sign at some radius and a repulsive force appears (that was first noted by \citet*{Gorkavyi} for
a simple quasi-Newtonian model). The repulsive force is very high at first moments of time then
it quickly decreases but never disappears.

A shape of the gravitational potential is often illustrated by a funnel made of rubber film
where a heavy ball is located in the centre. In this model a fast decrease of the gravitational mass
corresponds sharp ascent of the ball. The film attached to the ball forms a cone-type hill 
in the centre of the funnel. Light balls on the central cone run away from the centre.
The central cone expands fast but keeps it exterior slope; this corresponds to long-term repulsive force. 

Recently, there is a debate about back-reaction of small-scale inhomogeneities (e.g., galaxies, stars, black holes) which can contribute to the background metric (\citet*{Green}, \citet*{Buchert}). \citet{Buchert} considered the backreaction or effect of the formation of nonlinear structures on the expansion rate and spatial curvature as a possibility that could explain cosmological observations of acceleration of the Universe without the introduction of dark energy or modified gravity. According to \citet{Green} the effect of inhomogeneities on average properties of the modern Universe is merely due to gravitational radiation (though small) and contributes positive pressure which cannot act as dark energy. We demonstrate that the mergers of inhomogeneities like black holes, resulting in emissions of gravitational waves, can generate a repulsive gravitational force. This result seems to contradict the Green and Wald theorem. These mergers act as an effective dark energy, if the total mass of the universe is decreased.

Let us consider a model of the universe with cycling periods of compression and expansion.
Gravitational waves caused by the compression may not disappear at the stage of expansion
and form the relic gravitational radiation. Because the gravitational radiation
does not contribute to the total gravitational mass of the Universe, a level of the energy
of the relic radiation is not limited and may be very high. According to \citet{Grishchuk} a maximum of
the energy of the relic gravitational radiation is at GHz frequencies. \citet{Zeldovich}
estimated a wavelength of the relic gravitational radiation be equal to $1.5*10^{-4}$ m.
This wavelength corresponds to a frequency of hundreds GHz.

\section{Conclusions}

We showed that a cosmic repulsive force can be explained in the classic general theory of relativity
without additional hypotheses. The repulsive force originates from a metric with
the varying gravitational mass of a system. The repulsive force occurs at some distances 
from the quasi-spherical system which depend on time lapsed from the beginning of the change
of the mass. The repulsive force quickly decreases with radius but does not disappear.

We hope that our theoretical prediction about decreasing acceleration of the Universe can be verified by observations.

It is logical to suppose that the found mechanism of the repulsive force may be applied to a model 
of the expanding universe. This may imply that big bang and accelerated expansion of the Universe
is not related to current processes in the Universe but to a relic repulsive gravitational force
or to a configuration of space-time that originates in the previous cycle of the Universe
when at the last stage of a collapse the intensive generation of gravitational waves
resulted in sharp decrease of the gravitational mass of the Universe (and may be in avoiding a singularity).
This process generated a powerful repulsive force that transformed big crunch into
big bang. At the early stage of big bang the repulsive acceleration was extremely high
(see the dropping branches of the curves 1 and 2 in Fig.~\ref{fig:fg2}).
Because the repulsive acceleration decreases with time, the current Universe expands
with lower acceleration (see the gently sloping parts of curve 3 in Fig.~\ref{fig:fg2}).
In a more realistic model, the parameter $\alpha$ may also depend on time.

The proposed metric with the varying gravitational mass of a system may be used
for the development of a cosmological model that explains the current expansion of the Universe
without assumptions of new fields and particles. Such a cosmological model may allow
an explanation of the anisotropy of movement of galaxies discovered by \citet{Kashlinsky}.

\section*{Acknowledgments}

The authors thank Alexander Kashlinsky, Sergei Kopeikin, Michail Ivanov, Igor Tkachev and a reviewer for helpful discussions and comments.












\bsp	
\label{lastpage}
\end{document}